\documentclass[aps,pra,twocolumn,letterpaper,superscriptaddress,10pt]{revtex4-2}

\usepackage{graphicx}
\usepackage{subfigure}
\usepackage{dcolumn}
\usepackage{bm}
\usepackage{subeqnarray}
\usepackage{xcolor} %
\usepackage{esvect}
\usepackage{ulem}
\usepackage{amsmath, amsthm, amssymb, amsfonts}
\usepackage{mathrsfs}
\usepackage{multirow}
\usepackage[colorlinks=true,urlcolor=blue,citecolor=blue,linkcolor=blue]{hyperref}

\begin{document}

\title{Quantum optical coherence: From linear to nonlinear interferometers}

\author{Kai-Hong Luo}
    \email{khluo@mail.uni-paderborn.de}
    \affiliation{Integrated Quantum Optics Group, Institute for Photonic Quantum Systems (PhoQS), Paderborn University, Warburger Stra\ss{}e 100, 33098 Paderborn, Germany}
\author{Matteo Santandrea}
    \affiliation{Integrated Quantum Optics Group, Institute for Photonic Quantum Systems (PhoQS), Paderborn University, Warburger Stra\ss{}e 100, 33098 Paderborn, Germany}
\author{Michael Stefszky}
    \affiliation{Integrated Quantum Optics Group, Institute for Photonic Quantum Systems (PhoQS), Paderborn University, Warburger Stra\ss{}e 100, 33098 Paderborn, Germany}
\author{Jan Sperling}
    \affiliation{Integrated Quantum Optics Group, Institute for Photonic Quantum Systems (PhoQS), Paderborn University, Warburger Stra\ss{}e 100, 33098 Paderborn, Germany}
\author{Marcello Massaro}
    \affiliation{Integrated Quantum Optics Group, Institute for Photonic Quantum Systems (PhoQS), Paderborn University, Warburger Stra\ss{}e 100, 33098 Paderborn, Germany}
\author{Alessandro Ferreri}
    \affiliation{Department of Physics and CeOPP, Paderborn University, Warburger Strasse~100, 33098 Paderborn, Germany}
\author{Polina R. Sharapova}
    \affiliation{Department of Physics and CeOPP, Paderborn University, Warburger Strasse~100, 33098 Paderborn, Germany}
\author{Harald Herrmann}
    \affiliation{Integrated Quantum Optics Group, Institute for Photonic Quantum Systems (PhoQS), Paderborn University, Warburger Stra\ss{}e 100, 33098 Paderborn, Germany}
\author{Christine Silberhorn}
    \affiliation{Integrated Quantum Optics Group, Institute for Photonic Quantum Systems (PhoQS), Paderborn University, Warburger Stra\ss{}e 100, 33098 Paderborn, Germany}

\begin{abstract}
    Interferometers provide a highly sensitive means to investigate and exploit the coherence properties of light in metrology applications.
    However, interferometers come in various forms and exploit different properties of the optical states within.
    In this paper, we introduce a classification scheme that characterizes any interferometer based on the number of involved nonlinear elements by studying their influence on single-photon and photon-pair states.
    Several examples of specific interferometers from these more general classes are discussed, and the theory describing the expected first-order and second-order coherence measurements for single-photon and single-photon-pair input states is summarized and compared. 
    These theoretical predictions are then tested in an innovative experimental setup that is easily able to switch between implementing an interferometer consisting of only one or two nonlinear elements.
    The resulting singles and coincidence rates are measured in both configurations and the results are seen to fit well with the presented theory. 
    The measured results of coherence are tied back to the presented classification scheme, revealing that our experimental design can be useful in gaining insight into the properties of the various interferometeric setups containing different degrees of nonlinearity.
\end{abstract}

\maketitle

\section{Introduction}
    
    Although the nature of photons has been investigated for many decades, new insights into their fundamental properties, such as coherence and wave-particle duality, continue to be developed \cite{TonomuraJOP1989, PanRMP2012multiphoton,AspdenAJP2016, JacksonAJP2018, PursehousePRL2019,Hochrainer2021quantum}. 
    Many of these findings, in both the classical and quantum regime, have been obtained via the use of optical interferometers, which provide an ideal platform for the investigation of the coherence properties of light \cite{HBT1956,AasiNP2013,LemosN2014quantum,SparaciariPRA2016,SlussarenkoNP2017,TanRP2019}.
   To probe the coherence properties of photons, both single-photon \cite{GrangierEL1986,BraigAPB2003} and two-photon interference \cite{HOMPRL1987,ShihPRL1988,RarityPRA1990two} have been explored in common interferometer setups, such as the Mach-Zehnder interferometer (MZI) and Hong-Ou-Mandel (HOM) interferometer configurations.
    However, investigations into  nonlinear interferometers, a subset of which are commonly labeled as SU(1,1) interferometers \cite{YurkePRA1986,ZouPRL1991induced}, have shown that these systems can also be used to characterize coherence properties of the quantum light involved and may even offer practical quantum advantages \cite{ChekhovaAOP216,CavesAQT2020,OuAPLP2020}.
    Recent experiments employing this class of interferometers have shown compelling results for both fundamental and applied physics, such as the recent demonstrations of wide-field interferometry and biphoton shaping \cite{OnoOL2019observation,FrascellaOptica2019,RiaziQI2019,PaterovaLSA2020,Vanselow20}.

    Owing to these recent results, a more complete understanding of the role of first-order classical  and higher-order quantum coherence in these systems is needed.
    In this paper, we explore the coherence properties of various interferometric schemes through click measurements of the output fields. 
    Interferometer configurations spanning from linear to nonlinear are considered and the measured data closely follows theoretical predictions.
    To obtain the pure investigation of coherence in different interferometers, here we only consider vacuum-seeded nonlinear processes, since the coherence property dramatically changes with the seeded input field.
    
    The paper is structured as follows:
    A consistent framework for comparing the various interferometeric schemes is first introduced in Sec. \ref{interferometers}.
    Two instructive examples in linear interferometery are then reviewed in Sec. \ref{lin-interferometers}, before the theory describing the expected fringing patterns for both the singles and coincidence clicks of the output beams for both seminonlinear (Sec.  \ref{semiinterferometers}) and nonlinear interferometers (Sec. \ref{nonlin-interferometers}) is established. 
    Extending upon previous theory, a model for the nonlinear interferometer including the effects arising due to broadband radiation is developed and discussed in Sec. \ref{nonlin-interferometers}.
    Section \ref{4-interferometers} then provides a summary and comparison of the various interferometer geometries. 
    Next, in Sec. \ref{experiment}, we present results from an experimental setup designed in such a way that minor modifications to the setup allows for investigation of both seminonlinear and nonlinear interferometer configurations.
    Finally, in Sec. \ref{result}, the experimental results are discussed and compared with theoretical predictions.

\section{Interferometers}
\label{interferometers}

\begin{figure}[tb]
    \center
    \includegraphics[width=\columnwidth]{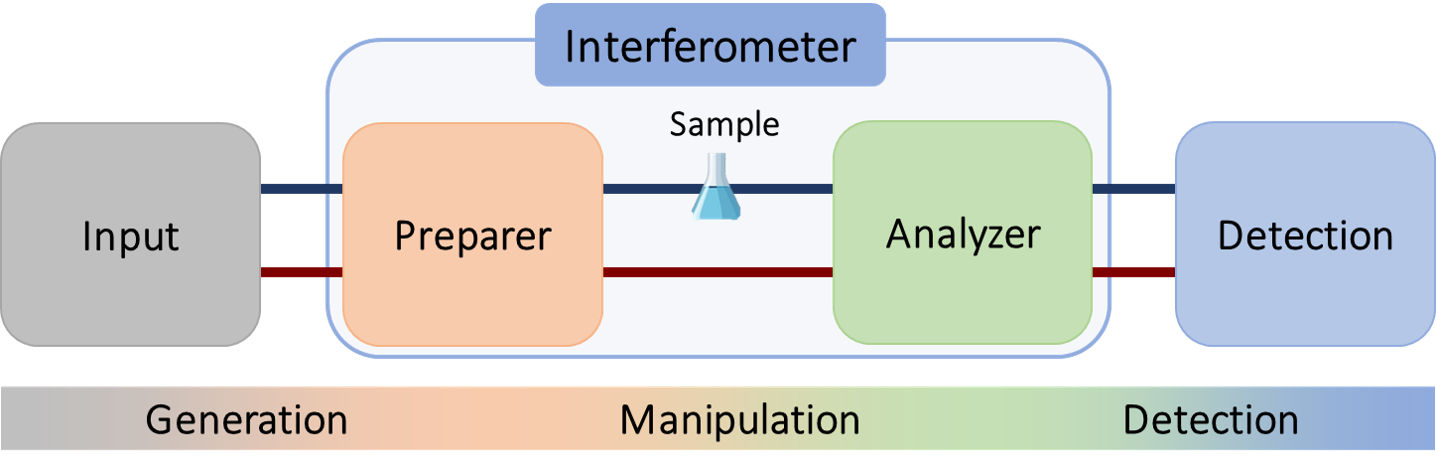}
    \caption{\label{Interferometer_General}
        General scheme of an optical interferometric setup, showing that the setup can be decomposed into its constituent parts. 
        The decomposition into generation, manipulation, and detection stages for many systems is somewhat arbitrary and is explored in this work.
    }
\end{figure}

    Any quantum optics experiment can be subdivided into three stages: a generation stage, a manipulation stage, and a detection stage. 
    Naturally, this subdivision can also be applied to all interferometric systems. 
    This concept is illustrated in Fig. \ref{Interferometer_General}.
    Within the interferometer itself, one can also identify three distinct elements: the preparer, the sample, and the analyzer.
    Typically, the sample can be considered as a differential phase or a temporal delay added to one of the beam paths between preparer and analyzer. 
    
    In this paper, we consider the case of a two-mode interferometer and define the input as the two modes entering the preparer. 
    The two modes exiting the analyzer are detected to make a measurement. Furthermore, we exclusively consider the case in which the two modes interact in the preparer and analyzer elements.
    Within this framework, an interferometer can then be classified into further classes by noting the characteristics of the preparer and the analyzer;
    depending on the physics involved in the mixing of the two input modes, each element can be defined as either a linear or a nonlinear interaction.
    
    An example of a linear element is constituted by beam splitters (BSs) while examples of nonlinear elements are three- and four-wave mixers.
    An interferometer that consists of only linear mixing elements, we denote as a linear interferometer, and one that consists of only nonlinear mixing elements is henceforth referred to as a nonlinear interferometer. 
    A system then consisting of one nonlinear stage and one linear stage is referred to as a seminonlinear interferometer, presenting an intermediate, hybrid interferometer configuration.

    The subdivision into the three stages---generation, manipulation, and detection---in an interferometric setup is, in some ways, arbitrary. 
    For example, it is not uncommon to consider the preparer as part of the state preparation and the analyzer as part of the detection scheme \cite{SparaciariPRA2016, WangPRA1991induced, ZouPRL1992observation, GraysonPRA1993interference, ZouPRA1993control}.
    For reasons that are later elucidated and that arise more critically in the case of nonlinear elements within the interferometer, arguably the most natural ways to define the various elements are as follows:
    the generation stage encompasses the elements after which we can say with certainty that our photons have been produced,
    while the detection stage is considered as any components situated behind the analyzer, such as detectors.
    Any components not involved in the photon generation and the detection then comprise the manipulation stage.

    The behavior of these interferometeric systems is further dependent on the chosen input state and detection stage. 
    For the input state, one can consider, for example, coherent states, single-photon states, and even vacuum.
    For the detection stage, one can consider intensity monitoring and homodyne setups \cite{WisemanPRL1993quantum, IbnoussinaOL2020heterodyne}, single-photon detection schemes, such as temporal bucket detection which is defined as averaging of the photon flux over the measurement time \cite{DevrelisAO1995,BraigAPB2003} and photon-number-resolved detection \cite{AchillesJMO2004photon}, and also more exotic schemes, such as coincidence \cite{HBT1956,HerzogPRL1994} and parity measurements \cite{ZhangPRA2019nonlinear}. 

\begin{figure*}
    \center
    \includegraphics[width=\textwidth]{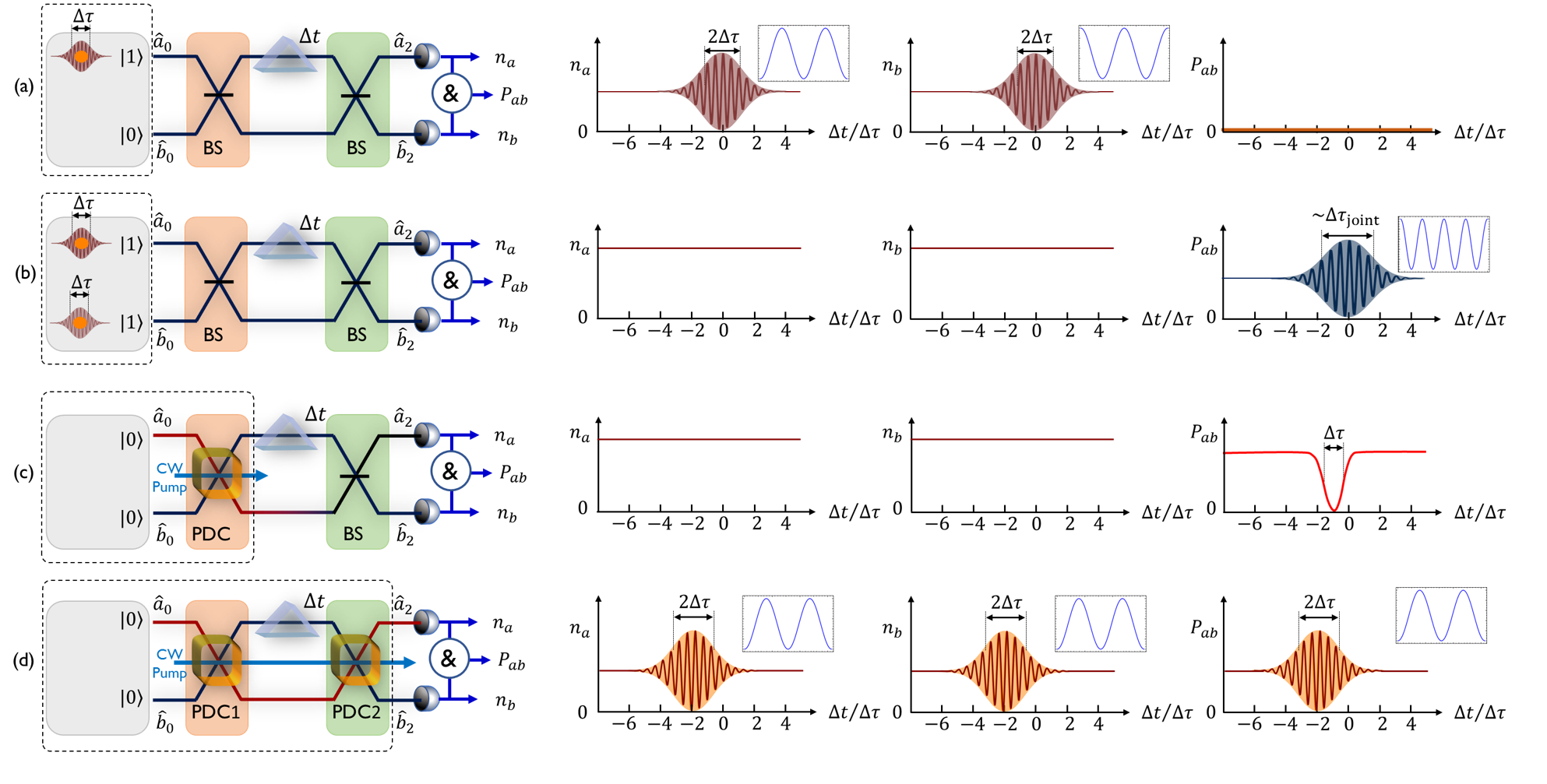}
    \caption{\label{Interferometer_Principle}
        Schematic of various interferometeric systems. 
        (a) Single-photon input state in a linear interferometer.
        (b) Single-photon pair input state in a linear interferometer, which may exhibit entanglement.
        (c) Vacuum-seeded seminonlinear interferometer, producing identical signal and idler photons from a parametric down-conversion (PDC) process.
        (d) Vacuum-seeded nonlinear interferometer with PDC stages as the nonlinear elements.
        For each interferometer configuration, both mean photon number from each output, $n_a$ and $n_b$, and coincidence probability, $P_{ab}$, are illustrated to the right of the setup. 
        The inserts show details of the fringing pattern around the region of maximum interference.
        The dashed box indicates what is considered as the generation stage in each optical interferometer.}
\end{figure*}

    An exhaustive review of all possible input states and detection stages is not feasible.
    Instead, we carefully choose a number of instructive cases and use these to demonstrate a number of key features for the three interferometer classes, i.e., linear, seminonlinear, nonlinear interferometer.
    In all cases, lossless interferometers are considered and both single-output bucket detection and coincidence detection schemes are analyzed to probe both classical and quantum coherence properties of these systems. 
    The considered input states are either single photons, allowing for entanglement between multiple photons, or vacuum.
    A summary of our selected interferometeric systems with corresponding input states, detection schemes, and predicted measurement results are shown in Fig. \ref{Interferometer_Principle}.

\subsection{Linear interferometer}
\label{lin-interferometers}

    We begin the discussion of the coherence properties of the chosen interferometers by reviewing linear interferometers with one- and two-photon input states, i.e., cases (a) and (b) in Fig. \ref{Interferometer_Principle}. 
    More specifically, we consider a typical MZI comprising two BSs acting as both preparer and analyzer, as depicted in Figs. \ref{Interferometer_Principle}(a) and \ref{Interferometer_Principle}(b).
    Note that this setup describes a transformation that lies within the SU(2) group and is therefore often called an SU(2) interferometer.
    As emphasized before, the subdivision into state generation, manipulation, and detection is in many ways arbitrary \cite{SparaciariPRA2016}.
    For example, one can consider that the inputs are transformed at the preparer into a new state that forms the arms of the interferometer. 
    By interfering these two modes on the analyzer and detecting the intensity of the outputs, one then performs a standard first-order cross-correlation measurement between these two modes.

\subsubsection{Linear interferometer with a single photon as input}
\label{1lin}

    We first consider the case of the linear interferometer seeded by a single photon in input mode $a$, $\left|1_a, 0_b\right\rangle$. 
    In this case, the generation of the photon occurs before entering the interferometer and as such the interferometer fulfills the role of manipulation of the input state.
    The photon is characterized by a well-defined spectral amplitude $A(\omega)$ and can be written as
    \begin{equation}
        \left|1_a, 0_b\right\rangle = \int d\omega A(\omega)\hat{a}^\dag(\omega)\left|\mathrm{vac}\right\rangle,
    \end{equation}
    where the normalization condition for $A$ reads $\int d\omega |A(\omega)|^2=1$.
    
    When an adjustable delay $\Delta t$ is inserted into one arm of the interferometer, the mean photon number $n_{a,b}$ from each output $a,b$, of the linear interferometer can be described via
    \begin{eqnarray}
    \label{eq:MZI_Sin}
        {\left\langle {{n_{a,b}}} \right\rangle}_{\rm{LI}}(\Delta t) 
        & = & \int_{-\infty}^{\infty} d{\omega} \left|A (\omega)\right|^2 \left|1 \pm \exp\left[ {i \left({\omega} \Delta t \right )}\right]\right|^2,
    \end{eqnarray}
    where the index LI is used to label results from this linear interferometer.
    With the help of the Wiener-Khinchin theorem \cite{Kubo2012statistical,LeibovichPRL2015aging}, the normalized expected mean photon number in each arm from Eq. \eqref{eq:MZI_Sin} can be rewritten as
    \begin{eqnarray}
    \label{eq:MZI_Sin_Cal}
        {\left\langle {{n_{a,b}}} \right\rangle}_{\rm{LI}}(\Delta t) 
        & = & \frac{1}{2} \mp \frac{1}{2} \left| g^{(1)}(\Delta t) \right| \cos \left({\omega_{0}} \Delta t \right ),
    \end{eqnarray}
    where ${\omega_{0}}$ is the central frequency  of the injected photon and $g^{(1)}$ is the degree of first-order temporal coherence of the input light, defined as $g^{(1)}(\Delta t)=G^{(1)}(\Delta t)/G^{(1)}(0)$.
    Here, $G^{(1)}(\Delta t)$ is the inverse Fourier transform of the input intensity spectrum, i.e.,  $G^{(1)}(\Delta t)=\mathcal{FT}^{-1}\left [\left|A(\omega)\right|^2 \right]$.
    The mean photon number from one output of this interferometer can be used to determine the first-order degree of coherence $g^{(1)}(\Delta t)$ of the input photon.
    The interference pattern is characterized by a fringing of period $2\pi/\omega_0$, and its envelope is given by $\left|g^{(1)}(\Delta t)\right|$.

    Having access to both output modes of the interferometer, it is possible to also measure the coincidences between them.
    However, for the considered case of a perfect single photon, no coincidences can be measured, i.e.,
    \begin{eqnarray}
    \label{eq:MZI_Coin_Cal}
        {\left\langle {P_{ab}} \right\rangle}_{\rm{LI}}(\Delta t) & = & 0, 
    \end{eqnarray}
    as illustrated in the rightmost plot of Fig. \ref{Interferometer_Principle}(a). 

    For the nontrivial case, let us consider here the case of a single photon with Fourier-transform-limited  amplitude profile, specifically, the Gaussian profile
    \begin{eqnarray}
    \label{eq:Gauss_Aw}
        A(\omega) = \frac{1}{\sqrt[4] {2\pi \sigma^2}} \exp{\left[-\frac{(\omega-\omega_0)^2}{4\sigma^2}\right]},
    \end{eqnarray}
    where $2\sigma$ is the $1/e^2$ spectral width.
    This is linked to the full width half maximum (FWHM) of the input photon's temporal duration $\Delta\tau$ by $\sigma= \sqrt{2\ln(2)}\left( \Delta \tau \right)^{-1}$.
    Such a photon is characterized by a $g^{(1)}(\Delta t)$ that is given by
    \begin{eqnarray}
    \label{eq:Gauss_g1}
        g^{(1)}(\Delta t) = \exp{\left[- \ln(2)
        \frac{\Delta t^2}{ \Delta \tau^2} \right]},
    \end{eqnarray}
    which has a FWHM of $2 \Delta \tau$.
    Note that the maximum visibility occurs at zero time delay, i.e., $\Delta t=0$, corresponding to equal path lengths for the two arms within the interferometer.
    Furthermore, there is a $\pi$-phase shift between the interference fringes in the two singles measurements.
    This behavior is a consequence of the fact that summing the two outputs for any delay necessarily gives a total mean photon number of one, i.e., $\left\langle n_a+n_b \right\rangle=1$.

\subsubsection{Linear interferometer with single-photon pair input}
\label{2lin}

    Now, we consider the scenario of a photon pair, $\left| {1_a, 1_b} \right\rangle $, acting as the input state, as pictured in Fig. \ref{Interferometer_Principle}(b). 
    As in the previous case, photon generation occurs before state manipulation through the interferometer, and we allow for the possibility of spectral entanglement between these two photons.
    This input state can be described in terms of the joint spectral amplitude (JSA) $J(\omega_a, \omega_b)$, i.e.,
    \begin{equation}
        \label{eq:11input}
        \left| {1_a, 1_b} \right\rangle = \iint d\omega_a d\omega_b J(\omega_a, \omega_b) \hat{a}^\dagger(\omega_a)\hat{b}^\dagger(\omega_b)\left| \mathrm{vac} \right\rangle.
    \end{equation}
    The mean number photon for both outputs are 
    \begin{eqnarray}
    \label{eq:QI_Sin_Cal}
        {\left\langle {{n_a}} \right\rangle}_{\rm{QI}}(\Delta t) =
        {\left\langle {{n_b}} \right\rangle}_{\rm{QI}}(\Delta t) = 1,
    \end{eqnarray}
    where the label QI indicates the linear interferometer with the single-photon pair input quantum state.
    Equation \eqref{eq:QI_Sin_Cal} shows that measuring the singles cannot provide any information about the two-photon state at the input of the interferometer.
    This is further illustrated in the two leftmost graphs in Fig. \ref{Interferometer_Principle}(b).

    In contrast, the coincidence pattern between the output modes is influenced by the JSA of the two-photon state, and thus partial information about the input state can be retrieved from the correlation measurement.
    The general expression for the coincidence probability, as a function of the time delay inside the interferometer, is quite complex and its envelope crucially depends on the shape of the JSA. 
    One can show that, in the case of identical input photons centered at $\omega_0$, the fringing seen in the coincidences is given by $\cos \left(2{\omega_{0}} \Delta t \right )$, which may lead to a quantum advantage in metrology applications, i.e., which may allow for determination of an unknown phase beyond what is classically achievable \cite{CamposPRA1990,SlussarenkoNP2017}.
    It can be also shown that it is possible to retrieve some information about the input two-photon states, such as their joint coherence time $\Delta \tau_\mathrm{joint}$, from the properties of the observed fringing in the coincidences \cite{CamposPRA1990,BishtJOSAB2015}.
    
    For the special case of indistinguishable photons, where the two photons have identical marginals, a path-entangled NOON state, $(|2,0\rangle+|0,2\rangle)/\sqrt 2$, is generated inside the interferometer.
    This state has been explored in detail in the context of quantum interferometry and metrology \cite{MitchellN2004super,AfekS2010high}. 
    In this case, the interference pattern is given by Ref. \cite{CamposPRA1990}
    \begin{equation}
    \label{eq:QI_Coin_Cal}
    \begin{aligned}
        {\left\langle {P_{ab}} \right\rangle}_{\rm{QI}}(\Delta t) 
        =  \frac{1}{2} {+}  \frac{1}{2} \exp{\left[-\frac{\Delta t^2}{\Delta \tau_\mathrm{joint}^2}\right]}\cos \left( 2{\omega_{0}} \Delta t \right ),
    \end{aligned}
    \end{equation}
    where $\omega_{0}$ is the central frequency 
    of the two photons, and $\Delta\tau_\mathrm{joint}$ is the width along the $+45^\circ$ bisector of the joint temporal intensity of the two-photon state, defined in Ref. \cite{CamposPRA1990}. 
    The derivation and analysis of Eq. \eqref{eq:QI_Coin_Cal} is beyond the scope of this paper; further details can be found in Ref. \cite{CamposPRA1990}.
    This interference pattern is illustrated in the rightmost plot of Fig. \ref{Interferometer_Principle}.
    Here, the mean number photon at both interferometer outputs is exactly the mean number of single-photon pairs at the input for all delays, i.e., $\left\langle n_a \right\rangle= \left\langle n_b \right\rangle=1$.
    Note that the point of maximum visibility occurs when the two arms of the interferometer are equal, as is the case for the single-photon input.
    The fact that double fringing is only observed in the coincidences, despite being constant in the singles, indicates the presence of higher-order, likewise quantum, coherence in the system.

\subsection{Seminonlinear interferometer}
\label{semiinterferometers}

    Next, we consider the seminonlinear interferometer, where either the preparer or the analyzer stages are nonlinear mixing elements, constituting the first nontrivial deviation from a purely linear configuration studied previously.

    Here, we specifically investigate the properties of one particular interferometer, in which the preparer is a nonlinear three-wave mixing stage, and the analyzer is a linear BS. 
    In contrast to the previous two cases, photon generation in this interferometer can only be said to have occurred after the preparer.
    For simplicity, we assume that only a single-photon pair is generated by the nonlinear parametric down-conversion (PDC) process. 
    Under these conditions, the resulting two-photon interference is the well-known HOM inteference effect \cite{HOMPRL1987,ShihPRL1988,RarityPRA1990two,ScheelPRA2003}.
    This seminonlinear interferometer setup is illustrated in Fig. \ref{Interferometer_Principle}(c).

    To provide a theoretical description of the process, we assume that the signal and idler photons are generated via a type-II, frequency-degenerate PDC process with continuous-wave (cw) pumping.
    The polarization of one field is rotated within the interferometer such that both photons interfere at the BS, the analyzer.
    Note that it is necessary to employ here a type-II process to ensure the separation of the output radiation in two different modes, while it is important to have frequency-degenerate signal and idler photons to ensure interference.

    With these considerations, the mean number photon at the output of the interferometer can be shown to obey
    \begin{eqnarray}
        {\left\langle {{n_a}} \right\rangle}_{\rm{HOM}}(\Delta t) =
        {\left\langle {{n_b}} \right\rangle}_{\rm{HOM}}(\Delta t) = 1.
    \end{eqnarray}
    Similar to the linear interferometer with two single photons, as previously discussed in Sec. \ref{2lin}, information about the generated photon structure and the interferometer delay cannot be gained via measurements of the mean number photon.

    Nevertheless, some information about the photon structure can be revealed by second-order measurements.
    For simplicity, let us consider a Gaussian approximation for the PDC process, including the potential presence of spectral filters  (see Appendix \ref{appendix_NonGer}). The JSA of the biphoton state is then given by    \begin{eqnarray}
    \label{eq:Gauss_PDC_process}
        \phi(\Omega) = \frac{1}{\sqrt[4] {2\pi \sigma^2}} \exp{\left[-\frac{\Omega^2}{4\sigma^2}\right]}\exp{\left[-i\Delta\tau_{0} \Omega\right]},
    \end{eqnarray}
    where $\Omega =\omega_s-\omega_{s0} = \omega_{i0}-\omega_i$ is the detuning of each photon from their central frequency, $\sigma = \sqrt{2\ln(2)}\left( \Delta \tau \right)^{-1}$ is again related to the bandwidth of either photons, and $\Delta \tau_{0}$ is the temporal walk-off acquired by the two photons when generated inside the nonlinear stage. 
    In this case, the coincidences between the two outputs of such an interferometer are given as \cite{CamposPRA1990,Walmsley1997}
    \begin{equation}
    \label{eq:HOM_Coin_Cal}
    \begin{aligned}
        &{\left\langle {P_{ab}} \right\rangle}_{\rm{HOM}}(\Delta t) \\
        = & \frac{1}{2}-\frac{1}{2} \int_{-\infty}^{\infty} d{\Omega} \left|\phi (\Omega)\right|^2 \exp \left[{ i 2 \Omega (\Delta t-\Delta\tau_0) }\right]
        \\
        = & \frac{1}{2}-\frac{1}{2} \exp \left[ {- \ln (2)\frac{(\Delta t-\Delta\tau_{0})^2}{ \left(\Delta\tau /2 \right)^2}} \right].
    \end{aligned}
    \end{equation}
    It shows that the width of the HOM dip is proportional to the photon's temporal bandwidth $\Delta \tau$.
    The position of the minimum in this dip is related to the temporal walk-off $ \Delta\tau_{0}$ of the generated biphoton state, the sign of which is determined by which photon is delayed.
 
    This seminonlinear interferometer shows a number of interesting features. 
    As was the case for the linear interferometer with single-photon pair input, one observes that the mean photon numbers are constant.
    The mean number photon is independent of the delay, and the mean number is given by the average nonlinear photon-pair generation rate, i.e., $\left\langle n_a \right\rangle= \left\langle n_b \right\rangle=1$.
    Similarly, the coincidence probability does vary as the arm delay in the interferometer is varied, but the resulting profile, the well-studied HOM dip, is very different. 
    The minimum of the HOM dip is found at $\Delta t = -\Delta \tau_0$, corresponding to the situation where the generated photons arrive at the analyzer BS simultaneously.
    It has been shown in a previous work \cite{Walmsley1997} that both the width of the HOM dip and visibility are related to the exchange symmetry of the generated photons.
    
    Another kind of seminonlinear interferometer consists of a BS and PDC element as the preparer and analyzer,  respectively.
    In such a scenario, and depending on the input state, the analyzer can be seeded with path-entangled light, resulting in yet another interesting form of coherence that is, however, not discussed in the context of this paper.

\subsection{Nonlinear interferometer}
\label{nonlin-interferometers}

    Finally, one can use nonlinear active processes as both the preparer and the analyzer, as shown in Fig. \ref{Interferometer_Principle}(d), resulting in what constitutes a nonlinear interferometer.
    In this paper, we consider the case of an interferometer composed of two identical, second-order nonlinear three-wave mixing processes.
    Such a system can be described by a set of Bogoliubov transformations, belonging to the SU(1,1) group \cite{YurkePRA1986}. For this reason, this subset of nonlinear interferometers are often referred to as SU(1,1) interferometers.
    These interferometers have been the focus of much investigation as it has been shown that they are able to provide quantum advantage for a number of applications, such as metrology \cite{ChekhovaAOP216,CavesAQT2020,OuAPLP2020}.

    To simplify the discussion of such second-order nonlinear interferometers, we consider the special case where the pump has a sufficiently low power and is a cw field such that, at most, only one photon pair is generated in the interferometer.
    The first fringing properties of cw-pumped second-order nonlinear interferometers have been explored in Ref. \cite{HerzogPRL1994}.
    However, this work did not fully consider the influence of the envelope of the spectrum of the generated light on the interference.
    Here, we expand on previous theory \cite{Ferreri2021Quantumspectrally} by explicitly including the effect of the spectral properties of the generated photon pairs in a way that allows comparison between the considered interferometric setups.

    The two-photon state at the output of the interferometer can be written as the superposition of generating a photon pair in either one of the two identical nonlinear media, 
    i.e.,
    \begin{equation}
    \label{eq:SU11_state}
        \left| \Psi \right\rangle \approx \iint d\omega_s d\omega_i J_N(\omega_s, \omega_i) \hat{a}^\dagger(\omega_s)\hat{b}^\dagger(\omega_i)\left| \text{vac} \right\rangle,
    \end{equation}
    with 
    \begin{equation}
    \label{eq:PM-SU11}
    \begin{aligned}
        & {J_{\rm{N}}\left( {{\omega _s}{\rm{,}}{\omega _i}} \right)} \\
        =& J\left( {{\omega _s}{\rm{,}}{\omega _i}} \right)
        \Big [ 1 + \exp\big(i \left [\Delta k (\omega_s, \omega_i) L\right]
        \\
        &
        +i\left[\Phi + \Delta(\omega_s,\omega_i) \right ]\big) \Big].
    \end{aligned}
    \end{equation}
    Here, $J(\omega_s, \omega_i)$ is the JSA of the photon pair that is generated in either of the two nonlinear processes, $\Delta k$ is the phase mismatch of the nonlinear PDC process, $L$ is the length of the nonlinear waveguide that implements the PDC, $\Phi$ is the relative phase between the two nonlinear stages, and $\Delta$ is the phase shift accumulated by the photon pair between the two stages.
    Considering, for simplicity, a time delay $\Delta t_s$ in the signal arm, this phase shift can be written as $\Delta = \Delta t_s\omega_s$.
    Please notice that, despite appearing very similar, Eq. \eqref{eq:11input} describes the two-photon input state of the interferometer, while Eq. \eqref{eq:SU11_state} describes the output of the nonlinear interferometer under the low-efficiency approximation.
    When the system is pumped with cw light at $\omega_{p0}$, the JSA $J_N(\omega_s, \omega_i)$ coincides with the phase-matching spectrum of the process, which is given by
    \begin{equation}
    \label{eq:PM-SU11_CW}
    \begin{aligned}
     & 
     {\phi_{\rm{N}}} {({\omega_s}\rm{,}{{\omega_{p0}}-{\omega_s}})} \\ 
    = & \phi ({{\omega _s}{\rm{,}}{\omega_{p0}-\omega _s}} )
     \Big[ 1 + \exp\big(i \left [\Delta k (\omega_s, \omega_{p0}-\omega _s) L\right]
    \\
    &
     + i\left[\Delta t_s \omega_s +\Phi \right ]\big) \Big].
    \end{aligned}
    \end{equation}

    From Eq. \eqref{eq:SU11_state}, the singles and the coincidence probability can be calculated as
    \begin{equation}
    \label{eq:SU11_Coin_Cal}
    \begin{aligned}
        & {\left\langle {n_{a}} \right\rangle}_{\rm{NI}}(\Delta t) =  {\left\langle {n_{b}} \right\rangle}_{\rm{NI}}(\Delta t) ={\left\langle {P_{ab}} \right\rangle}_{\rm{NI}}(\Delta t)
        \\
        = & \int_{-\infty}^{\infty} d{\omega} \left|\phi_N (\omega)\right|^2
        \\
        = &  {\frac{1}{2}} + {\frac{1}{2}} \left| h^{(1,1)} \left( \Delta t \right) \right| \cos \left({\omega_{s0}\Delta t +\Phi}\right),
    \end{aligned}
    \end{equation}
    resulting in Fig. \ref{Interferometer_Principle}(d). 
    Here, $\omega_{s0}$ is the central frequency of the signal photon.
    The interference pattern in Eq. \eqref{eq:SU11_Coin_Cal} has an envelope $\left| h^{(1,1)} \left(\Delta t \right) \right|$, which is proportional to the twin beam correlation function ${G^{(1,1)}(\Delta t)}$, and the period of fringing is decided by the wavelength of the photon that experiences the time delay.

    Once again, we approximate the PDC process using the Gaussian profile to simplify the treatment.
    Under this assumption, the envelope of the nonlinear interference pattern is given by
    \begin{equation}
    \label{eq:SU11_g11}
    \begin{aligned}
        h^{(1,1)} \left( \Delta t \right) 
        =\exp{\left[- \ln(2)
        \frac{ \left(\Delta t -2\Delta\tau_{0} \right)^2}{ \Delta \tau^2} \right]}.
    \end{aligned}
    \end{equation}
    This equation reveals that the width of the interference envelope of both the singles and the coincidences is given by $2\Delta \tau$.
    Moreover, it shows that the position of maximum visibility occurs at a delay of $2\Delta\tau_{0}$.
    Note that the delay of $2\Delta\tau_{0}$ is equivalent to the walk-off exhibited by signal and idler photons generated in a sample of twice the length $2L$.
    
    One can observe that the same interference pattern characterizes both the singles and the coincidence probability, in stark contrast to all previously explored systems.
    This is due to the fact that the singles in both arms oscillate in phase, according to the equations that govern the evolution of the state in such an interferometer. 
    In particular, for a lossless nonlinear interferometer, one finds that  $\left\langle n_a \right\rangle= \left\langle n_b \right\rangle=\left\langle P_{ab} \right\rangle$.
    Therefore, the (first-order) measurement of the singles provides the same information that the (second-order) coincidence measurements would provide, as previously highlighted in Ref. \cite{ZouPRL1991induced}.

\subsection{Comparison among various interferometers}
\label{4-interferometers}

    A comparison of the interference patterns presented by the interferometers in Fig. \ref{Interferometer_Principle} reveals a key concept within these interferometers that follow our classification in terms of the degree of nonlinearity.
    In particular, the observed interference patterns are related to the impossibility of retrieving information about the evolution of the state within the interferometer.
    In the presented linear and seminonlinear setups, this is due to erasure of the which-way information of the different paths taken by the state leading to the detectors \cite{Hochrainer2021quantum}.
    In contrast, in the nonlinear interferometer, the coherent generation of a single pair of photons from the interferometer precludes knowledge of which-stage information; i.e., one cannot say at which stage, preparer, or analyzer, the photon pair has been generated. 
    This reveals that optimal interference between the two stages will only occur when the state generated in the preparer is indistinguishable in all degrees of freedom to the state generated in the analyzer. 
    It is for this reason that it is necessary to consider the entire nonlinear interferometer as the region of state generation in this scenario.
    
    An additional, detailed overview over the aforementioned interferometers with their input states is presented in Appendix \ref{4-inter}.

\begin{figure*}[bt]
    \center
    \includegraphics[width=0.95\textwidth]{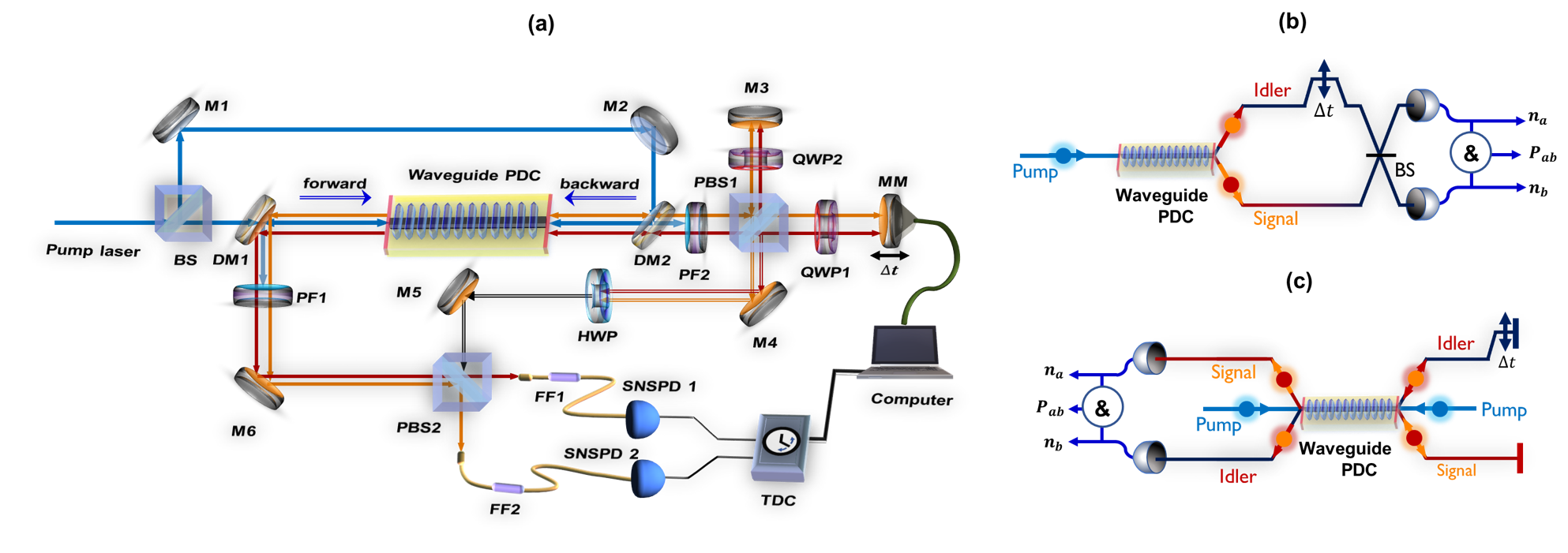}
    \caption{\label{WaveguidePDC2_SU11Setup}
        Experimental setup (a) and corresponding conceptual overview of seminonlinear (b) and folded, nonlinear interferometer (c) based around a single nonlinear waveguide. 
        Full details of the setup are provided in the text. 
        The delay stage used in both experimental geometries is identical.
        (b) Illustration of a seminonlinear interferometer.
        When the reflected pump beam in the backward direction is blocked, i.e., only the forward pump is present, the QWP is turned to 45$^\circ$, and the HWP is turned to 22.5$^\circ$, we have the HOM interference occurring at PBS2.
        (c) Schematic diagram of folded nonlinear interferometer.
        When the QWP is set at 0$^\circ$, we directly and individually count the number of signal and idler photons because of the polarization-dependent splitting.}
\end{figure*}

\section{Experimental realization}
\label{experiment}

    To experimentally investigate the theoretical work thus far presented, a setup that can readily switch between seminonlinear and nonlinear interferometer configurations was constructed.
    The concept is that of a folded geometry design with a waveguide PDC source is utilized as the nonlinear medium, as depicted in Fig. \ref{WaveguidePDC2_SU11Setup}. 
    A single pass through this device produces signal and idler photons which can then be interfered on a BS to realize the seminonlinear interferomter configuration, as illustrated in Fig. \ref{WaveguidePDC2_SU11Setup}(b).
    The nonlinear interferometer is then implemented by retro-reflecting the generated signal and idler photons back through the waveguide after some free-space propagation. 
    Traversing the same waveguide twice ensures that the phase-matching profiles of the preparer and analyzer stages are indeed identical.

    The detailed experimental setup is depicted in Fig. \ref{WaveguidePDC2_SU11Setup}(a). 
    A 19-mm-long titanium indiffused lithium niobate waveguide serves as the nonlinear medium, with a periodic poling period of 9.3~$\mu$m chosen for type-II PDC at around 160~$^{\circ}$C, with a stability of approximately $\pm$5~mK. 
    Both end-facets of the waveguide, as well as all coupling lenses, have anti-reflection coatings for the telecom wavelength range.
    The waveguide is pumped by a cw external cavity diode laser, working at a wavelength of 777~nm and being separated into a forward and reverse propagating pump field through the use of a BS.  
    In this scheme, the forward propagating pump then drives the preparer and the reverse propagating pump field drives the analyzer.

    After each stage (preparer, analyzer), the signal and idler modes are separated from the pump laser by a dichroic mirror (DM1, DM2), and a coated silicon filter is then used to suppress the remaining pump (PF1, PF2). 
    
    The two orthogonally polarized forward-propagating signal and idler modes are separated by the polarizing beam splitters (PBSs).
    This separation allows for the insertion of a variable time delay between these modes via adjustment of the position of a movable mirror (MM), mounted on a computer-controlled translation stage.
    Two quarter-wave plates (QWPs) positioned between two high reflectivity mirrors (M3 and MM) and PBS1 are used to either direct the photons directly towards PBS2 (QWPs set to 45$^\circ$), where they interfere with the aid of the half-wave plate (HWP set to 22.5$^\circ$) to realize a seminonlinear interferometer or the photons are directed back through the waveguide (QWPs set to 0$^\circ$) to realize a nonlinear interferometer. 
    Note that, in the case of the seminonlinear interferometer, the reverse-propagating pump is blocked. 

    The fields at the two output ports of PBS2 are coupled to single-mode fibers. 
    To suppress background photons, each fiber is connected to a $\sim$1.0~nm wide bandpass fiber filter (FF1 and FF2).
    The bandwidth of this filter is chosen to be slightly narrower than the expected bandwidth ($\sim$1.3~nm) of the photons generated in a single pass of the sample. 
    Finally, the photons are detected using superconducting nanowire detectors (SNSPDs) with a detection efficiency of around 90\% and a time-to-digital converter. 
    The details of the loss characterization and time referencing of the setup are given in Appendix \ref{ExpCha-setup}. 

\begin{figure*}[bt]
    \centering
    \includegraphics[width=0.95\textwidth]{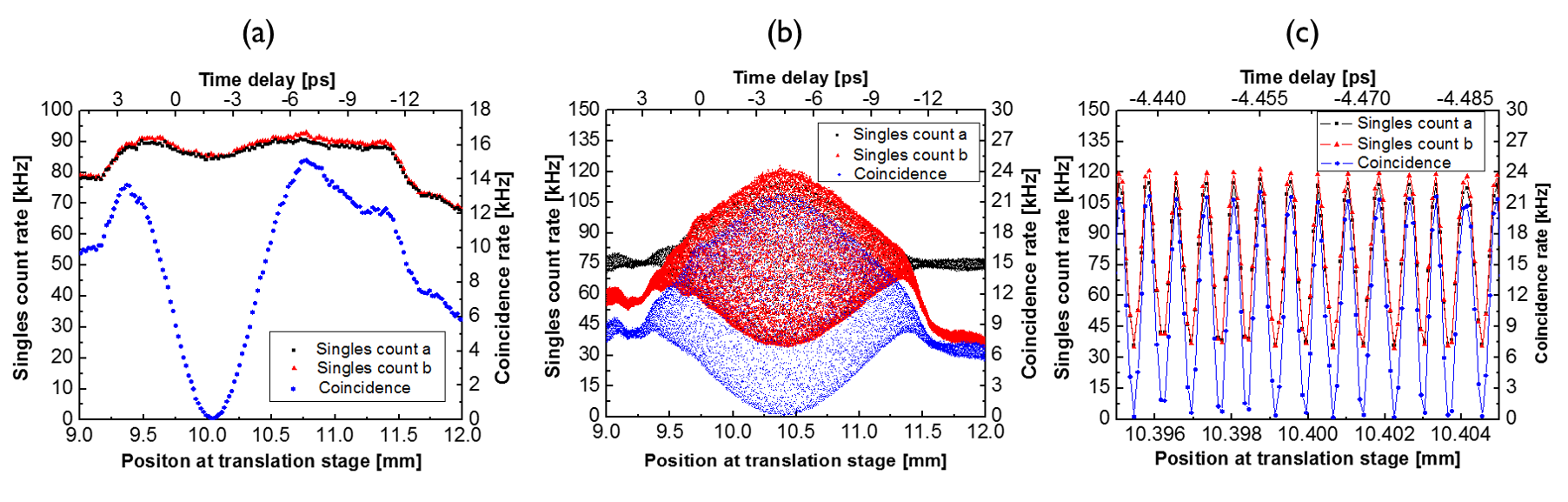}
    \caption{\label{PDC2_SU11_Exp01}
        Measured singles (black and red for photons $a$ and $b$, respectively) and coincidence (blue) count rates for seminonlinear (a) and nonlinear (b), (c) interferometer configurations as a function of the varied position of the movable mirror (MM) from Fig. \ref{WaveguidePDC2_SU11Setup}.
        Tick labels on top are the corresponding time delays.
        The reduction in count rates in (a) and (b) at larger translation stage positions are caused by a slight misalignment of the stage.
        Graph (c) shows a magnified view of the interference fringes at the time delay corresponding to maximum visibility ($-$4.50~ps). 
        Connecting solid lines are provided as a guide for the eye.
    }
\end{figure*}

\section{Experimental Results}
\label{result}

    The experimental setup is first configured to implement the seminonlinear interferometer, as described in the previous section.
    The delay, i.e., position of MM, is chosen such that the faster, vertically polarized idler photon experiences a shorter beam path than the other slower, horizontally polarized signal photon.
    With the help of HWP, two photons with orthogonal polarizations then arrive at PBS2 at the same time.
    At a pump power of $\sim$1~mW, a HOM dip with a visibility of (99.3$\pm$0.3)\% is measured for a stage position around 10.03~mm, corresponding to $\Delta \tau_{0}=-2.25$~ps in the time domain.
    This delay matches the expected temporal walk-off between the two photons when taking into account dispersion properties of our 19~mm long lithium niobate waveguide, as expected from Eq. \eqref{eq:HOM_Coin_Cal}.
    Note that the width of the seminonlinear HOM dip is determined by the spectral width of the detected photons, which, in the presented experimental setup, is affected by the chosen filters.
    The effect of these filters is explained in detail in Appendices \ref{appendix_NonGer} and \ref{appendix_Filter}.

    Next, the setup is then reconfigured as a nonlinear inteferometer by simply setting the QWPs to 0$^\circ$ and pumping the waveguide in both the forward and reverse directions.
    The ratio of the pump powers driving the preparer and analyzer are chosen such that they maximize the visibility of the coincidence counts \cite{OnoOL2019observation}.

    The measured singles and coincidence count rates are shown in Fig. \ref{PDC2_SU11_Exp01}(b) as the position of the MM is scanned over several millimeters with a resolution of $\sim$100~nm. 
    It is immediately apparent that the envelope of the observed interference traces for all three measurements is practically identical.
    The shape of these envelopes is affected by the spectral filtering, which is explained in detail in Appendix \ref{appendix_Filter}. 
    Moreover, by zooming into the region of highest visibility, one can see that all three traces exhibit identical frequencies and phases, Fig. \ref{PDC2_SU11_Exp01}(c).
    However, one should also note that the maximum visibility of the two singles count rates ($\sim$ 50\%) and the coincidence count rates ($\sim$ 100\%) vary significantly.
    This difference is mainly caused by imperfect coupling into waveguide and fibers as well as other losses in the setup.

    The delay required to reach the point of optimal visibility reveals further information about the presented system.
    As expected, the optimal visibility for the singles and coincidence count rates is found around a mirror position of 10.40~mm, corresponding to a time delay of $2 \Delta \tau_0 = -4.50$~ps; see Eq. \eqref{eq:SU11_g11}.
    This delay corresponds to twice the temporal walk-off ($\Delta \tau_{0}$) that one would expect between signal and idler photons when generating these photons in a single pass of the waveguide.
    This configuration corresponds to the situation where a photon pair generated in the preparer has the same delay between signal and idler photons as a photon pair generated in the analyzer upon detection.
    This ensures indistinguishability between generation in the first and second stages, a necessary condition to achieve perfect interference in the nonlinear interferometer \cite{HerzogPRL1994}.

\begin{figure}[bt]
    \center
    \includegraphics[width=\columnwidth]{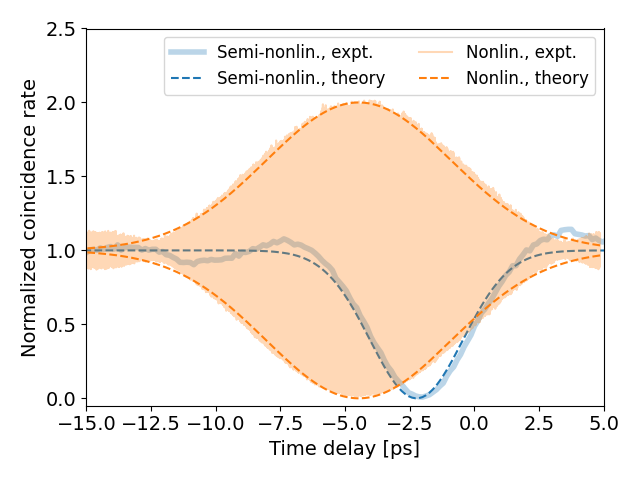}
    \caption{\label{theVSexp}
        Comparison of theory and measured results from the reconfigurable experimental setup for a 19-mm-long sample with 1-nm-broad spectral filters.
        The experiment is first configured for seminonlinear HOM-type interference (cyan), before being reconfigured to realize a nonlinear interferometer (orange). 
        Solid lines correspond to the measured experimental results while 
        dashed lines correspond to the theoretical envelopes given by Eqs. \eqref{eq:HOM_Coin_Cal} and \eqref{eq:SU11_Coin_Cal}.
    }
\end{figure}

    Finally, Fig. \ref{theVSexp} provides a comparison between the theory in Eq. \eqref{eq:SU11_g11} with the measured coincidence envelopes, shown in Fig. \ref{PDC2_SU11_Exp01}, for both the seminonlinear and nonlinear experimental configurations.
    One can see that the theory precisely describes the measured behavior, highlighting both the stability of the experimental setup and the validity of the presented theory.
    The effect of the frequency filtering on the output spectrum has to be taken into account and is presented in detail in Appendix \ref{appendix_Filter}.

\section{Conclusions}
\label{conclusion}

    In summary, we introduced a meaningful classification system for interferometers according to their degree of nonlinearity, and we carried out a comparative study of the impact of the degree of nonlinearity on coherence properties of quantum light.
    The degree of nonlinearity is particularly important when attempting to identify the generation, manipulation, and detection stages of any interferometeric setup.
    It was shown, for example, that one should consider the nonlinear interferometer as a single, albeit complicated, generation stage, as explored previously \cite{YurkePRA1986}.
    Such insight reveals that one would expect the singles and coincidence count rates to be identical, as was indeed measured, because of the pairwise generation that characterizes PDC systems.
    Furthermore, when optimizing the visibility in the nonlinear interferometer, one needs to consider that the interferometer delay should be set to maximize interference between the possibility of generating photons in either the preparer or the analyzer. 
    This means that the nonlinear interference perfectly happens when the temporal walk-off between two possible photons from the first preparer stage is elaborately compensated in the middle of the second analyzer stage.

    A number of instructive examples were chosen to highlight some of the coherence properties of interferometers within different classes. 
    The corresponding theory for each case was presented and the expected singles and coincidence number were described.
    Previous theory was expanded upon to accurately describe the results of experimental measurements for the case of the nonlinear interferometer.

    To experimentally investigate the theoretical predictions, an innovative experimental design was developed.
    Through simple rotation of a pair of wave plates, it is possible to configure the device as either a seminonlinear interferometer or a nonlinear one.
    The folded geometry ensures that the preparer and the analyzer stages are as identical as possible. 
    Furthermore, this experimental setup exploits a nonlinear waveguide as the core component, thereby increasing the strength of the nonlinear interaction and opening up the possibility for more complicated future devices in integrated architectures \cite{LuoSA2019}.

    The measurements taken on this setup were observed to very closely follow the expected theoretical behavior.
    This was seen in both the shape (envelope) and position (perfect constrictive and destructive interference) of the measured interference patterns in both seminonlinear and nonlinear configurations.
    These results highlighted the suitability of the presented experimental setup for jointly investigating and comparing different types of interferometers.
    The presented discussion and classification system shows the value in considering the similarities and dissimilarities of different interferometeric setups.
    This presents a step toward exploiting advanced nonlinear interferometry for applications, such as quantum imaging \cite{LemosN2014quantum}, sensing \cite{Simon2017quantum}, and quantum metrology \cite{Hochrainer2021quantum,PezzeRMP2018quantum}.

\begin{acknowledgments}
    We thank Vahid Ansari, Benjamin Brecht, Christof Eigner, and Raimund Ricken for helpful discussions and technical assistance.
    The Integrated Quantum Optics group acknowledges financial support through the Deutsche Forschungsgemeinschaft (DFG -- German Research Foundation) (NO. 231447078 -- TRR 142 C02), and the Gottfried Wilhelm Leibniz-Preis (Grant No. SI1115/3-1).
\end{acknowledgments}


\appendix* 
\section{}
\label{app}

\subsection{Comparison among various interferometers}
\label{4-inter}

    Table \ref{tab:4Inter_Result} is provided as a summary of the findings presented throughout the paper, including the information that is more conceptually given in Fig. \ref{Interferometer_Principle}.
    The figure provides the properties of the singles and coincidences measurements of all interferometer configurations presented in the paper.

\begin{table*}[hbt!] 
    \begin{center}
    \caption{
        Overview of interference properties of various interferometers with given input states. 
        All results assume that the interferometers are lossless, the spectra are well approximated by a Gaussian, and PDC processes are pumped by a cw laser.
    }\label{tab:4Inter_Result}
    \begin{tabular}{|c|c|c|c|c|c|}
      \hline
      \multicolumn{2}{|c|}{\multirow{3}{*}{\textbf{Interferometers}}}&  \multicolumn{2}{|c|}{\multirow{2}{*}{\textbf{Linear}}} & {\multirow{2}{*}{\textbf{Seminonlinear}}} & {\multirow{2}{*}{\textbf{Nonlinear}}}  \\
      \multicolumn{2}{|c|}{\multirow{3}{*}{}}&  \multicolumn{2}{|c|}{\multirow{2}{*}{(BS ans BS)}} & {\multirow{2}{*}{(PDC and BS)}} & {\multirow{2}{*}{(PDC and PDC)}}  \\ 
      \multicolumn{2}{|c|}{}&  \multicolumn{2}{|c|}{} & {} & {}\\
      \hline
      \multicolumn{2}{|c|}{\multirow{2}{*}{\textbf{Inputs}}}& {\multirow{2}{*}{$\left|1_a, 0_b\right\rangle$}}  & {\multirow{2}{*}{$\left|1_a, 1_b\right\rangle$}} & {$\left|0_a, 0_b\right\rangle$} &{$\left|0_a, 0_b\right\rangle$}\\ 
      \multicolumn{2}{|c|}{}& & {} & {plus cw pump}&  {plus cw pump}\\
      \hline
      {\multirow{4}{*}{\textbf{Singles}}} & {Envelope\phantom{$\Big|$}}  & { $ \left | g^{(1)}(\Delta t) \right |$ } & {\small{n/a}} & {\small{n/a}} &  {$ \left | h^{(1,1)}(\Delta t) \right |$}\\ \cline{2-6}
      {} & {FWHM\phantom{$\Big|$}}  & {$2 \Delta \tau$} & {\small{n/a}} & {\small{n/a}} & {$2 \Delta \tau$} \\  \cline{2-6}
      {} & {Offset\phantom{$\Big|$}}  & {$\Delta t =0$} & {\small{n/a}} & {\small{n/a}} & {$ \Delta t = \pm 2 \Delta \tau$}\\  \cline{2-6}
      {} & {Fringes\phantom{$\Big|$}}  & {$\mp \cos \left({\omega_{0}} \Delta t \right )$} & {\small{n/a}} & {\small{n/a}}  & {$\cos \left({\omega_{s0}\Delta t +\Phi_{\Lambda}}\right)$}\\ 
      \hline
      {\multirow{4}{*}{\textbf{Coincidence}}} & {Envelope\phantom{$\Big|$}}  & {\small{n/a}} & {Joint shape} & {HOM dip} & {$ \left | h^{(1,1)}(\Delta t) \right |$} \\ \cline{2-6}
      {} & {FWHM\phantom{$\Big|$}}  & {\small{n/a}} & {$\Delta \tau_\mathrm{joint}$} & {$\Delta \tau$} & {$2 \Delta \tau$} \\  \cline{2-6}
      {} & {Offset\phantom{$\Big|$}}  & {\small{n/a}} & {$\Delta t =0$} & {$ \Delta t = \pm \Delta \tau$} & {$ \Delta t = \pm 2 \Delta \tau$} \\  \cline{2-6}
      {} & {Fringes\phantom{$\Big|$}}  & {\small{n/a}} & {$\cos \left(2{\omega_{0}} \Delta t \right )$} & {\small{n/a}}  & {$\cos \left({\omega_{s0}\Delta t +\Phi_{\Lambda}}\right)$}\\
      \hline
      \multicolumn{2}{|c|}{\multirow{2}{*}{\textbf{Conservation}}}&\multicolumn{2}{|c|}{photon-number conservation} & \multicolumn{2}{|c|}{photon-pair conservation}\\ \multicolumn{2}{|c|}{}& \multicolumn{2}{|c|}{$n_a+n_b= n_\mathrm{input}$} & \multicolumn{2}{|c|} {$n_a=n_b= n_{pp}$}\\
     \hline
    \end{tabular}
    \end{center}
\end{table*}

\subsection{Experimental characterization of setup}
\label{ExpCha-setup}

    Before going into the detailed properties of two interferometers, the loss performance of the whole setup and the behavior of photon-pair generation via a single PDC process are characterized.
    First, we consider that there is only a backward pump launched from the rear side of the waveguide.
    According to the single count rates and coincidence count rate, the Klyshko efficiencies of both arms are about 21\% and 26\%, respectively.
    While the forward pump is injected from the front side of waveguide, both generated signal and idler photons are reflected and again pass through the waveguide from the rear.
    Because of increased coupling losses due to traveling through the same waveguide twice, the Klyshko efficiencies of both arms then drop to 6\% and 7\%, respectively.
    Therefore, the internal losses between two PDC processes, which means the out-coupling loss from the waveguide and in-coupling loss into the waveguide, are around 29\% and 27\%, which includes waveguide coupling, fiber coupling, and all other optical components in the setup. 
    
    A benefit of the folded geometry is that the absolute zero-point of the  time delay $\Delta t=0$ can be determined.
    In this way, the position of the interference fringes can be unambiguously determined without a priori assumptions, such as the dispersion of the nonlinear material.
    This calibration is achieved in two separate steps.
    First, the apparatus is configured as a seminonlinear interferometer.
    The crystal is pumped in the forward direction only (the reverse direction pump is blocked) and the two QWPs are set to 45$^\circ$, thereby avoiding traversing the nonlinear material a second time.
    Interference between signal and idler then occurs at PBS2.
    The resulting HOM dip position then corresponds to half of the crystal dispersion, i.e., $\Delta \tau_{0}/2$.  
    The second step consists of setting the QWPs to 0$^\circ$, thereby directing the beams back through the nonlinear material a second time, noting again that the reverse direction pump is blocked.
    The position of the HOM dip now measured corresponds to the usual factor of $\Delta \tau_{0}/2$ plus the dispersion added by traversing the entire sample in the reverse direction; i.e., the final position will be $\Delta \tau_{0}/2+\Delta \tau_{0}$.
    From these two positions, the zero time delay position $\Delta t =0$ can be deduced.
 
\subsection{Modelling of parametric downconversion in integrated nonlinear waveguides}
\label{appendix_NonGer}

    A PDC process in a periodically poled nonlinear waveguide is a widespread efficient method to generate a stream of photon pairs. 
    In such a waveguide PDC process, by exploiting strong $\chi^{(2)}$ nonlinearity and long interaction length $L$, a single pump photon is split into two photons of lower energy, named signal and idler, according to energy and momentum conservation.

    A simplified treatment of the PDC process considers the ideal case case of lossless propagation inside the nonlinear crystal and ignores higher-order photon-number contribution; i.e., it considers a low gain regime.
    Under these assumptions, the state of the photon pairs generated via PDC in a periodically poled waveguide with length $L$ and poling period $\Lambda$ can be described as
    \begin{eqnarray}
    \label{eq:PDCstate0}
    \begin{aligned}
        &{\left| \Psi  \right\rangle _{\rm{PDC}}} \\
        \propto &\int d{\omega _s}d{\omega _i}{f_L \left( {{\omega _s}{\rm{,}}{\omega _i}} \right)} \hat a_s^\dag \left( {{\omega _s}} \right)\hat a_i^\dag \left( {{\omega _i}} \right)|\mathrm{vac}\rangle,
    \end{aligned}
    \end{eqnarray}
    where ${{\hat a^\dag}_{s,i}}\left( {{\omega _{s,i}}} \right)$ describes the photon creation operator at frequency ${{\omega _{s,i}}}$ and  ${f_L \left( {{\omega _s}{\rm{,}}{\omega _i}} \right)} = \alpha \left( {{\omega _s},{\omega _i}} \right) \phi_L\left( \omega_s\rm{,}\omega_i \right)$ is the JSA of the nonlinear waveguide.
    The JSA is determined by the pump $\alpha \left( {{\omega _s},{\omega _i}} \right)$ and the phase matching $\phi_L\left( {{\omega _s}{\rm{,}}{\omega _i}} \right)$. 
    The phase-matching spectrum is determined by the momentum conservation among three fields in the waveguide and can be expressed as
    \begin{eqnarray}
    \label{eq:PDCstateJSA0}
    \begin{aligned}
        \phi_L\left( {{\omega _s}{\rm{,}}{\omega _i}} \right) = &{\mathrm{sinc}} \left[ {\Delta \beta \left( {{\omega _s}{\rm{,}}{\omega _i}} \right)\frac{L}{2}} \right]\\
        &\times\exp{ \left[ i{\Delta \beta \left( {{\omega _s}{\rm{,}}{\omega _i}} \right)\frac{L}{2}} \right]},
    \end{aligned}
    \end{eqnarray}
    where $\Delta \beta \left( {{\omega _s}{\rm{,}}{\omega _i}} \right) = \Delta k \left( {{\omega _s}{\rm{,}}{\omega _i}} \right) -\frac{2 \pi}{\Lambda}$. 
    Here, $\Delta k(\omega_s, \omega_i) = k_p(\omega_s + \omega_i) - k_s(\omega_s) - k_i(\omega_i) $ is the wave vector mismatch of the process, and $k_{p,s,i}$ are the wave vectors of the three fields inside the waveguide.
    
    For the special case of a PDC that is pumped by a cw pump, one has that $\omega_p = \omega_s + \omega_i$ holds true.
    Thus, it is possible to simplify the PDC state as 
    \begin{eqnarray}
    \label{eq:PDCstate1}
    \begin{aligned}
        &{\left| \Psi  \right\rangle _{\rm{PDC}}}\propto \int d{\Omega} \phi(\Omega) \hat a_s^\dag \left( {\Omega} \right)\hat a_i^\dag \left( -\Omega \right)|\mathrm{vac}\rangle,
    \end{aligned}
    \end{eqnarray}
    where the detuning $\Omega$ is defined as $\Omega = \omega_s - \omega_{s0}$.
    In this case, one can usually consider the first-order Taylor expansion of $\Delta k(\omega)$ at the central wavelength ${\omega_{s0}}$, and thus Eq. \eqref{eq:PDCstateJSA0} can be simplified as
    \begin{eqnarray}
    \label{eq:PDCstateJSA1}
    \begin{aligned}
        \phi_L\left( \Omega \right) = {\rm {sinc}} \left[ {\Delta\tau_0 \Omega} \right]\exp{ \left[ i \Delta\tau_0\Omega \right]},
    \end{aligned}
    \end{eqnarray}
    where $\Delta\tau_0$ is the average temporal delay between by the two signal and idler photons, when they exit from the nonlinear medium.
    This is given by
    \begin{eqnarray}
    \label{eq:walkoff_expression}
    \begin{aligned}
        \Delta\tau_0 &= \frac{L}{2}\left[ \left.\frac{\partial k_i}{\partial \omega}\right|_{\omega=\omega_p - \omega_{s0}} -\left.\frac{\partial k_s}{\partial \omega}\right|_{\omega=\omega_{s0}}\right]\\
        &= \frac{L}{2}\left[ \frac{1}{vg_i}-\frac{1}{vg_s}\right],
    \end{aligned}
    \end{eqnarray}
    where $vg_{s/i}$ are the group velocities of signal and idler photons.

    To simplify the calculations, it is useful to approximate Eq. \eqref{eq:PDCstateJSA1} with the Gaussian function
    \begin{eqnarray}
    \label{eq:GaussApproxPhasematching}
    \begin{aligned}
        \phi_L\left( \Omega \right) =\exp{\left[ -\gamma \Delta\tau_0^2 \Omega^2 - i \Delta\tau_0\Omega \right]},
    \end{aligned}
    \end{eqnarray}
    where $\gamma = 0.193$ is a coefficient used to match the amplitude bandwidths between the Gaussian approximation and the correct sinc spectrum.
    For the sample discussed in the main text, the expected $\Delta\tau_0$ is $\approx 2.2$~ps, given a sample length of $L\approx 19$~mm and the dispersion properties of lithium niobate at the wavelengths of interest.

\subsection{Spectral filter}\label{appendix_Filter}

    The fiber filters used in front of the detectors have been approximated by Gaussian functions, having a spectral amplitude of
    \begin{eqnarray}
    \label{eq:GaussFilters}
    \begin{aligned}
        f_g\left( \Omega \right) =\exp{\left[ -2\ln (2) \frac{\Omega^2}{\Delta\Omega^2}\right]},
    \end{aligned}
    \end{eqnarray}
    where $\Delta\Omega$ is the intensity FWHM of the filter, corresponding to 1~nm for the filters employed.
    Therefore, the effective JSA of the photon pair that is generated in the waveguide is given by
    \begin{equation}
    \label{eq:effectivePhasematching}
    \begin{aligned}
    \phi_{\rm{L}}(\Omega) =&f_g(\Omega)\phi_\text{L} (\Omega)\\
    =&\exp\left[ -  \frac{\Omega^2}{4\sigma^2}- i \Delta\tau_0\Omega \right],
    \end{aligned}
    \end{equation}
    with 
    \begin{equation} 
    \frac{1}{\sigma^2} =4\gamma \Delta\tau_0^2  +  \frac{8\ln(2) }{\Delta\Omega^2}.
    \end{equation}
    Considering a sample length of 19~mm and filters with an intensity FWHM of 1~nm, $\kappa$ is equal to $\sim$1.8~ps.
    From the effective JSA in Eq. \eqref{eq:effectivePhasematching}, one can derive the envelope of the seminonlinear and nonlinear interference as 
    \begin{eqnarray}
        \label{eq:envelopes}
        \text{HOM}(\Delta t)&= \exp\left[-\frac{(\Delta t - \Delta \tau_0)^2}{2\kappa^2}\right],
        \\
        h^{(1,1)}(\Delta t) &= \exp\left[-\frac{(\Delta t - 2\Delta \tau_0)^2}{8\kappa^2}\right].
    \end{eqnarray}
    Therefore, the expected FWHM for the seminonlinear and nonlinear interference pattern are, respectively, 4.3~ps and 8.6~ps and match quite remarkably the measured values.


\bibliography{SU11QuantumCoherence}

\end{document}